\documentclass[twocolum]{afmc_art}
\usepackage{graphicx}
\usepackage{bm}
\usepackage{paralist}
\usepackage{mathrsfs}
\renewcommand{\vec}[1]{\mbox{\boldmath $ #1 $}}

\newcommand{\ii}{\mathrm i}
\newcommand{\ee}{\mathrm e}
\newcommand{\dprime}{{\prime\prime}}
\newcommand{\intz}{\int_0^{\infty}}

\begin{document}
%
%
% --- Title goes here. Do not for a line-breack unless absolutely necessary
\title{Generation and Breakdown of Surface Streaks in Wind-Driven Aqueous Flow}

% --- Authors and affiliations with brief address
\author{Asim \"Onder$^1$\corresp{asim.onder@gmail.com},~~Philip L.-F. Liu$^{1,2,3}$ and Wu-ting Tsai$^4$}
\affiliation{$^1$ Department of Civil and Environmental Engineering\\
 National University of Singapore, Singapore 117576, Singapore \\[5pt]
 $^2$ School of Civil and Environmental Engineering, \\
 Cornell University, Ithaca, NY 14850, USA \\[5pt]
 $^3$ Institute of Hydrological and Oceanic Sciences, \\ 
 National Central University, Jhongli, Taoyuan, 320, Taiwan \\[5pt]
 $^4$ Department of Engineering Science and Ocean Engineering, \\ 
 National Taiwan University, Taipei, 10617, Taiwan\\[5pt]
}

\maketitle

%In wind-driven free-surface boundary layers, these turbulent eddies can be generated already before the onset of the waves when shear-induced small-scale (O(cm)) streaks break down.

% --- Paper proper starts here
\section{Abstract}
A bypass transition scenario in a wind-stress driven aqueous flow is analysed using
 a temporally developing boundary layer model with accelerating surface drift velocity. %The parameters of the model are selected to mimic a  wave-tank experiment with a reference wind speed of $5\mbox{~m/s}$. 
 To study the boundary layer processes in isolation, a  flat free surface is adapted, which inhibits the initiation of waves. First, preferred initial perturbations to which the boundary layer is the most sensitive are identified using linear non-normal growth theory. These perturbations are arranged as streamwise-constant vortex pairs located adjacent to the free surface. Subsequently, direct numerical simulations are initialized with these optimal perturbations and streamwise  streaks are generated. High-speed streaks penetrate into deeper water layers and undergo sinuous instabilities reminiscent of the instabilities developing on low-speed streaks in wall-bounded flows. Streak instabilities induce lateral undulations at the free surface, which closely resemble the dye patterns before the onset of waves in wind-wave-tank experiments. The present analysis provides a theoretical background for these experimental observations. 

\keywords
free-surface boundary layer, bypass transition, transient growth

\section{Introduction}
Boundary layers below the sea surface are well known to mediate the transport of heat and soluble gases across the air-sea interface. A much less studied aspect of these boundary layers is their possible role in the generation of wind waves under mild winds. To that end, there is some evidence that the initiation of first visible wind waves closely follows the transition path in the subsurface aqueous boundary layer  \cite{caulliez98,shrira2005}. Wind-wave tank experiment by Caulliez et al.~\cite{caulliez98} showed that this transition path resembles the bypass route to transition in flat-plate boundary layers. First, longitudinal streaks form right beneath the free surface. After a certain fetch depending on the wind speed, the streaks undergo oscillations and break locally into turbulent spots. These turbulent spots spread downstream and induce V-shaped zones at the free surface, where wind waves grow ``explosively" and become visible to unaided eye. The water surface remains calm in the laminar regions outside these V-shaped zones. Further downstream, the intermittent patches of turbulence eventually spread over the whole spanwise extent of the wave tank and the waves cover the whole water surface. To date, the building steps of this boundary layer transition are not studied systematically. The present work is an effort in this direction.

Motivated by the weakness of the waves in the laminar and transitional regions, Tsai et al.~\cite{tsai2005numerical} studied a wind-stress driven boundary layer using direct numerical simulations with a flat free surface. The seeded particles in these simulations formed very similar streaky patterns to those in a former wind-wave tank experiment \cite{melville_shear_veron_1998}. The flat free surface effectively captures the essentials of the initial boundary layer processes in the wind-driven flow of water.  

Bypass transition occurs when the finite-amplitude ambient perturbations are processed by the boundary layer and amplified to considerable magnitudes. In this process, the initial perturbations are commonly assumed to be sufficiently small to be analysed in linear regimes, and nonlinearity is restricted to the final stages of the transition. This allows a mode-by-mode analysis, in which the whole parameter space can be easily searched for the most dangerous perturbations leading to maximum perturbation growth. In linearly stable flows, this growth is of transient nature and is driven by non-normal mechanisms \cite{Schmid:2014cd}.  In the present work, we adapt this idealised approach to study the bypass transition in the wind-driven aqueous flow. As the waves remain very small until the onset of turbulence, we follow the simulation in \cite{tsai2005numerical} and use a flat free surface.  First, we will linearise the problem to find an optimal initial condition inducing the maximum growth in a given time frame. Subsequently, this optimal initial condition will be evolved into nonlinear regimes using direct numerical simulation and breakdown to turbulence will be analysed.

\section{Flow configuration}
The temporally developing boundary layer model in \cite{melville_shear_veron_1998} is selected to study the wind-driven flow of water. In this model, the free surface linearly accelerates with a drift velocity $U_s=At$, where the constant acceleration parameter $A$ is linked to a reference wind speed. The flow is defined in a Cartesian coordinate system with $x$, $y$ and $z$ correspond to streamwise, spanwise, and vertical directions. The latter extends from the free surface downwards. Consequently, the laminar base flow is given by:
\begin{eqnarray}
\label{eq:UMom}
	\frac{\partial U}{\partial t}&=&\nu \frac{\partial^2 U}{\partial x^2}, \\
	U(z,0)&=&0,\\
	\label{eq:windBC}
	\frac{\partial u(0,t)}{\partial z}&=&2 \frac{A\sqrt t}{\sqrt{\pi\nu}}, \\
%		\frac{\partial v(0,t)}{\partial z}&=& A_s \left (2 \frac{A\sqrt t}{\sqrt{\pi\nu}}\right) \cos(\beta y); \\
%	w(0,t)&=&0; \\
	\label{eq:Uz}
	U(z\rightarrow \infty,0)&=&0,
\end{eqnarray}
where equation~(\ref{eq:windBC}) represents the driving wind stress acting at the free surface. Equations~(\ref{eq:UMom})--(\ref{eq:Uz}) have the following similarity solution
\begin{equation}
	U(\eta,t)=At\left[(1+2\eta^2)\mathrm{erfc}(\eta)-\frac{2}{\pi^{1/2}}\eta \ee^{-\eta^2}  \right ]
\end{equation}
where $\eta=z/2(\nu t)^{1/2}$. 

In this study, we set $A=0.01\mbox{~m/s}^2$, which corresponds to a reference wind speed $U_w=5$ m/s, cf. \cite{melville_shear_veron_1998,tsai2005numerical} for details. For this case, streamwise streaks became evident at around $t\approx 21$ s, and started to exhibit three-dimensional features at $t\approx23$ s, cf. figure~2 in \cite{melville_shear_veron_1998}. 

\section{Linear transient growth}\label{sec:IO}
%\cite{onder_liu_2020}, \cite{onder2020receptivity}

We introduce three-dimensional perturbations on the base flow $U$ and track their evolution. The temporal evolution of these perturbations is most conveniently studied using the Orr--Sommerfeld and Squire equations~\cite{Schmid:2014cd}, i.e., 
\begin{eqnarray}
\label{eq:OSS1}
\left [\left (\frac{\partial  }{\partial t}+\ii\alpha U-\nu\hat \Delta \right)\hat\Delta-\ii\alpha \frac{\partial^2 {U}}{\partial z^2}\right ]\hat w&= &0, \\ 
\label{eq:OSS2}
\left (\frac{\partial  }{\partial t}+\ii\alpha U-\nu\hat\Delta \right) \hat \eta+ \ii \beta \frac{\partial U}{\partial z}\hat w&=&0, \\
\hat w(z,t_o)=\hat w_o;~~~~ \hat \eta(z,t_o)&=&\hat \eta_o,\\
\hat w(0,t)=\hat \eta(0,t)&=&0, \\
\frac{\partial \hat w}{\partial z}(0,t)=-\ii\alpha\hat u(\hat w,\hat \eta)-\ii\beta\hat v(\hat w,\hat \eta),&& \\
\label{eq:OSS5}
\hat w(z\rightarrow\infty,t)=\frac{\partial \hat w}{\partial z}(z\rightarrow\infty,t)=\hat\eta(z\rightarrow\infty,t)&=&0, 
\end{eqnarray}
where $\hat\Delta=\partial^2/\partial z^2-k^2$ with $k^2=\alpha^2+\beta^2$, and $\hat w$ and $\hat \eta=\ii\beta\hat u-\ii\alpha \hat v$ are the Fourier modes of vertical velocity and vorticity:
\begin{equation}
\label{eq:uFourier}
[u^\prime,\eta^\prime](x,y,z,t)=\mathrm {Re}\{ [\hat u,\hat \eta]  \mathrm e^{\ii (\alpha x+ \beta y)}\}.
\end{equation}

We look for the optimal initial conditions at a wavenumber pair $(\alpha,\beta)$ seeded at a time $t_o$, which yields the strongest growth at a terminal time $t_f$ measured by the perturbation kinetic energy~\cite{Schmid:2001hj}
\begin{eqnarray}
E(\hat{\vec q}):=\frac{1}{2k^2}\intz (k^2|\hat w|^2+|\frac{\partial \hat w}{\partial z}|^2+|\hat \eta|^2)\mathrm dz.
\label{eq:E}
\end{eqnarray}
This is equivalent to
\begin{eqnarray}
\label{eq:GOpt}
G_f(\alpha,\beta;t_o,t_f,A)&:=&\max\limits_{ \vec {\hat q}_0}\frac{E(\vec {\hat q}(t_f))}{E(\vec {\hat q}_0)}, 
\end{eqnarray}
where $G_f$  is the maximum growth, $\hat {\vec q}_o=[\hat w_o, \hat \eta_o]$ and $\hat {\vec q}=[\hat w, \hat \eta]$. The optimization problem is solved using an adjoint approach~\cite{Luchini:2014fv}. To this end, the growth $G_f$ is maximum when the flow is initialized 
with
\begin{equation}
\hat w_o=-\frac{2k^2 E(\hat q(t_f))}{E(\vec {\hat q}_o)}\hat w^+(z,t_o),~~~\hat \eta_o=\frac{2k^2 E(\hat q(t_f))}{E(\vec {\hat q}_o)}\hat \eta^+(z,t_o),
\end{equation}
where $\hat w^+$ and $\hat \eta^+$ are the adjoint vertical velocity and vorticity fields satisfying 
the following adjoint Orr--Sommerfeld and Squire equations~\cite{Schmid:2001hj}
\begin{eqnarray}
\label{eq:adjOSS1}
\left [\left (\frac{\partial  }{\partial t}+\ii\alpha U+\nu\hat{\Delta} \right)\hat{\Delta}+2\ii\alpha \frac{\partial {U}}{\partial z}\frac{\partial}{\partial z}\right ]\hat w^+=-\ii \beta \frac{\partial U}{\partial z}\hat \eta^+, &&\\ 
\left (\frac{\partial  }{\partial t}+\ii\alpha +\nu\hat{\Delta} \right) \hat \eta^+= 0, && \label{eq:adjOSS2}\\
\hat w^+(z,t_f)=-\frac{1}{2k^2}\frac{\hat w(z,t_f)}{E(\vec {\hat q}_o)},~~~\hat \eta^+(z,t_f)=\frac{1}{2k^2}\frac{\hat \eta(z,t_f)}{E(\vec {\hat q}_o)},&&\\
\hat w^+(0,t)=\hat \eta^+(0,t)=0,&& \\
\frac{\partial \hat w^+}{\partial z}(0,t)=-\ii\alpha\hat u^+(\hat w,\hat \eta)-\ii\beta\hat v^+(\hat w,\hat \eta),&& \\
\label{eq:adjOSS5}
\hat w^+(z\rightarrow\infty,t)=\frac{\partial \hat w^+}{\partial z}(z\rightarrow\infty,t)=\hat\eta^+(z\rightarrow\infty,t)=0.&& 
\end{eqnarray} 
The forward and adjoint Orr--Sommerfeld and Squire equations are solved in an iterative fashion using an adjoint-looping algorithm \cite{Andersson:1999ej}. To this end, the equations are discretized using a spectral method based on Chebyshev polynomials \cite{Weideman:2000hq}. Converged results are obtained for a domain size $z\in[0,10\mbox{~cm}]$ and a resolution of $N_z=61$ Chebyshev collocation points in the vertical direction. The Crank--Nicolson scheme is employed for time integration.   

\begin{figure}[h!]
\centering
\includegraphics[]{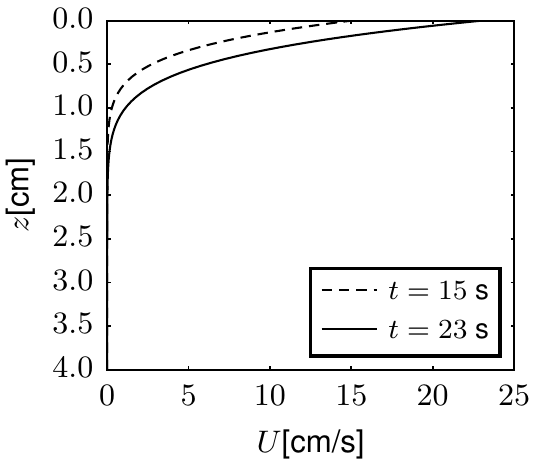}
\caption{\label{fig:UBase} Vertical profiles of the base flow at times $t=15$ s and $23$ s. }
\end{figure} 

\begin{figure}[h!]
\centering
\includegraphics[]{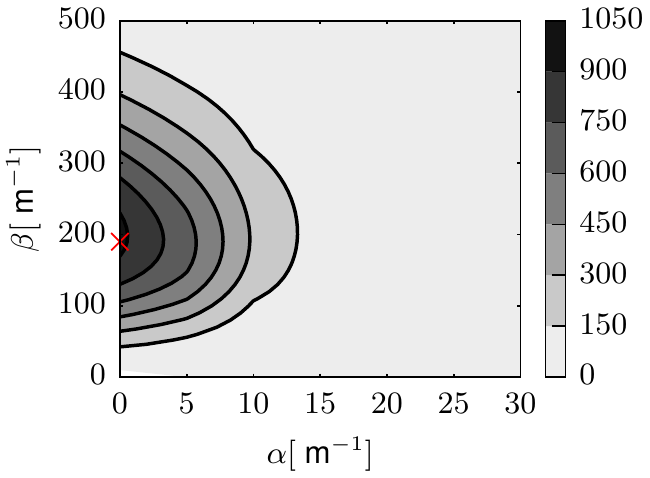}
\caption{\label{fig:Gmax} Contours show the maximum growth $G_f(\alpha=0,\beta=190\mbox{~m}^{-1}; 15~\mbox{s}, 23~\mbox{s}, 0.01\mbox{~m/s}^2)$  for each wavenumber pair.}
\end{figure}

We have selected $t_o=15$~s and $t_f=23$~s to mimic the physics in \cite{melville_shear_veron_1998}. The profiles of the base flow at these time instances  are plotted in figure~\ref{fig:UBase}. The maximum growth $G_f(\alpha,\beta; 15~\mbox{s}, 23~\mbox{s}, 0.01\mbox{~m/s}^2)$ for each wavenumber pair is demonstrated in figure~\ref{fig:Gmax}. Streamwise-constant modes ($\alpha=0,\beta\neq0$) clearly dominate with amplifications reaching $10^3$ at the range $\beta\approx 190-200~\mbox{m}^{-1}$. The streak spacing for the case with $U_w=5$~m/s is approximately 3.3~cm in \cite{melville_shear_veron_1998}. Neglecting the streamwise variations, and assuming periodicity in the spanwise direction, this spacing corresponds to wavenumber pair $(\alpha=0,\beta\approx 190~\mbox{m}^{-1})$ (shown with red marker in figure~\ref{fig:Gmax}). The streamwise-constant modes in our analysis are an approximation to these streaks, and we observe an excellent match between the wavenumbers of the most-amplified and experimental streaks.

\begin{figure}[h!]
\centering
\includegraphics[]{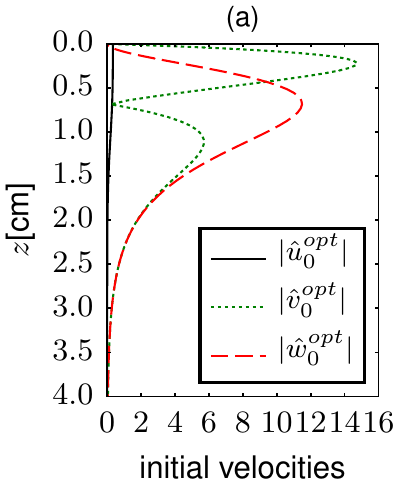}
\includegraphics[]{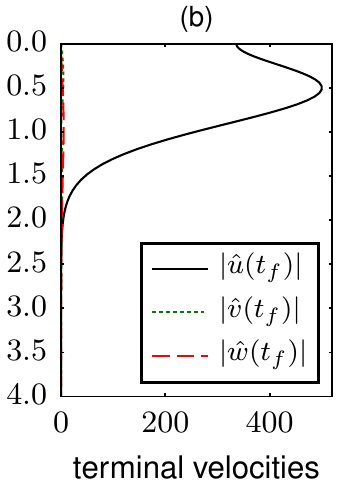}
\caption{\label{fig:u0} (a) Optimal initial condition calculated for $\alpha=0$, $\beta=190\mbox{~m}^{-1}$, $t_o=15\mbox{~s}$, and $t_f=23\mbox{~s}$ is demonstrated using amplitudes of velocity components. (b) The response of the flow at the terminal time for optimization ($t_f=23\mbox{~s}$). }
\end{figure}

The optimal initial condition delivering $G_f(\alpha=0,\beta=190\mbox{~m}^{-1}; 15~\mbox{s}, 23~\mbox{s}, 0.01\mbox{~m/s}^2)$ is plotted in figure~\ref{fig:u0}a using the amplitudes of Fourier modes in Cartesian coordinates. The streamwise component is vanishingly small compared to the cross-stream components. Figure~\ref{fig:u0}b further plots the velocities at the terminal time $t_f=23\mbox{~s}$. This time the streamwise component clearly dominates. It makes a peak at $z\approx0.5$ cm and does not vanish at the free surface. Concentration of driving perturbations in cross-stream components and the flow response in the streamwise component is typical for the transient non-normal growth processes in steady  \cite{Schmid:2014cd}  and unsteady \cite{onder_liu_2020} shear flows. Physically, cross-stream components form counter-rotating vortex pairs, which stir the boundary layer and amplify streamwise velocity component (streaks) \cite{landahlJFM80}, see next section. This mechanism is often called the lift-up effect.

\begin{figure}[h!]
\centering
\includegraphics[]{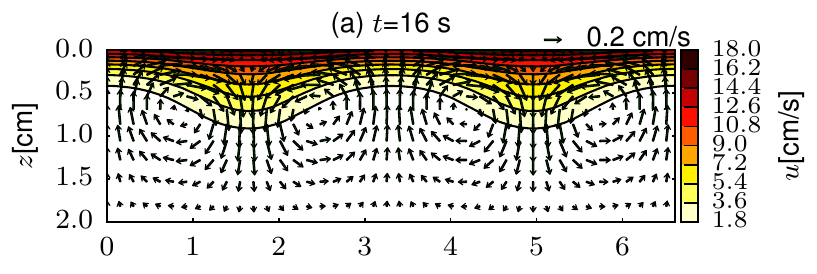}
\includegraphics[]{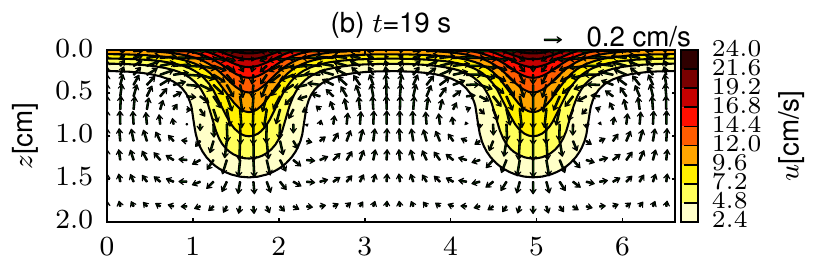}
\includegraphics[]{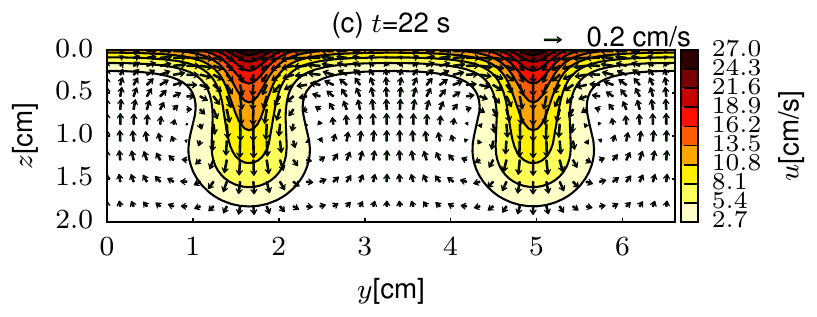}
\caption{\label{fig:UVW} Instantaneous fields at the vertical cutplane at $x=1~\mbox{cm}$. }
\end{figure}

\section{Breakdown to turbulence}
In the further stages of bypass transition nonlinearity should be accounted. To this end, we evolve the optimal initial condition, which is obtained for $\alpha=0$, $\beta=190\mbox{~m}^{-1}$, $t_o=15\mbox{~s}$, $t_f=23\mbox{~s}$ and $A=0.01\mbox{~m/s}^2$, with nonlinear Navier--Stokes equations. A mixed spatial discretization is utilized, in which bi-dimensional spectral-elements are employed in ($x-z$) plane, and Fourier extensions are applied in $y$ direction \cite{cantwell2015nektar++}.  The computational domain is a rectangular box with dimensions $13.22\mbox{~cm}\times 6.61\mbox{~cm}\times 10\mbox{~cm}$. A structured grid consisting $60 \times 40$ spectral elements is used in ($x-z$) plane, where the elements are clustered towards the free surface.  The spectral elements are  equipped with 6th order interpolation polynomials. In the spanwise direction, $96$ Fourier modes are defined. Periodic boundary conditions are employed in the streamwise and spanwise directions. At the free surface ($z=0$), the vertical velocity vanishes, and the wind-stress boundary condition in equation~(\ref{eq:windBC}) is imposed for the streamwise velocity. The free-slip boundary condition is applied at the bottom boundary. The amplitude of the streamwise-constant initial field is set to $2.5\times10^{-2}\mbox{~cm/s}$. A smaller amplitude ($2\times10^{-3}\mbox{~cm/s}$) white noise field is superposed on this initial field.

\begin{figure}[h!]
\centering
\includegraphics[width=78mm]{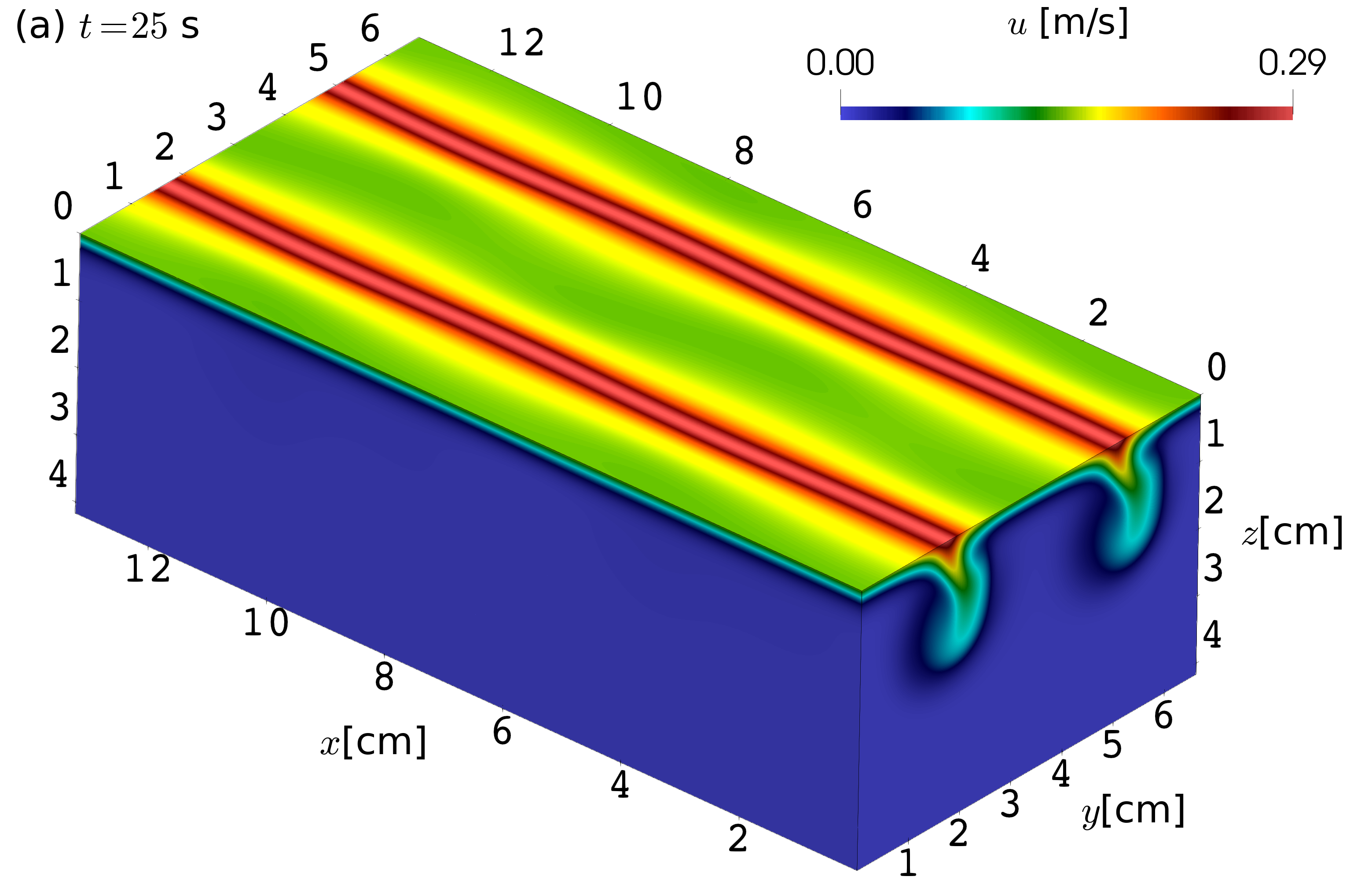}
\includegraphics[width=78mm]{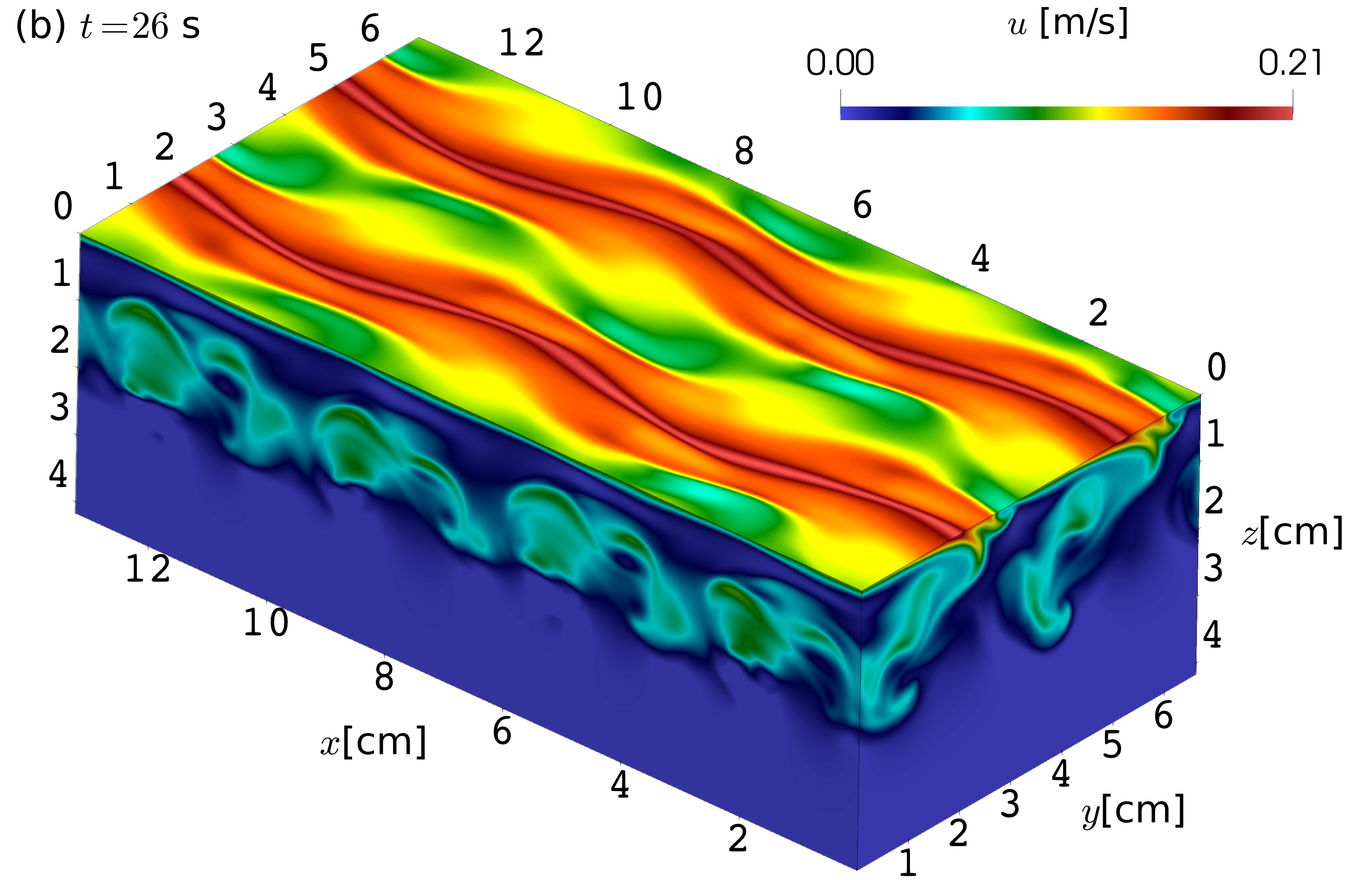}
\includegraphics[width=78mm]{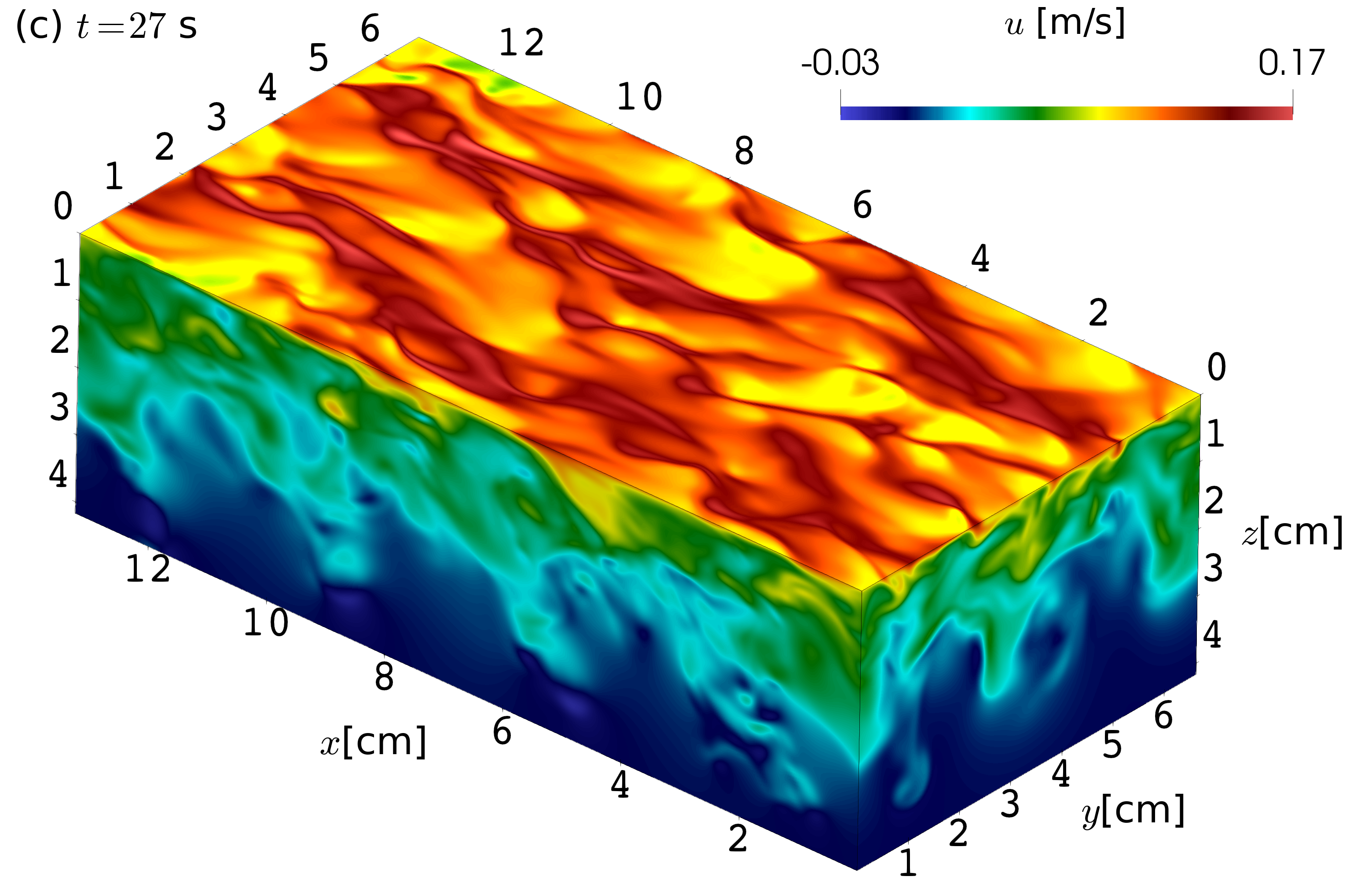}
\caption{\label{fig:viz3D} Breakdown to turbulence.}
\end{figure}

The development of the flow in the initial stages are plotted in figure~\ref{fig:UVW}.  Cross-stream components are arranged as counter-rotating vortex pairs, which eject high-speed fluid from the free-surface downwards. This mechanism produces high-speed streaks, which amplify in time and penetrate into the deeper layers of water.  The penetrated streaks possess spanwise and vertical shear layers and are vulnerable to inflectional instabilities. Consequently, a sinuous instability is observed, which manifests itself as meandering high-speed streaks at the free surface, cf. figure~\ref{fig:viz3D}b. These sinuous undulations at the free surface resemble the streak undulations in wind-wave tanks  \cite{caulliez98,shrira2005}. Following these instabilities, the streaks rapidly break down and a fully turbulent boundary layer sets in at $t=27\mbox{~s}$ (figure~\ref{fig:viz3D}c). The onset of turbulence yields significant reduction in streamwise momentum, which is consistent with the experiments, cf. e.g. figure~3a in \cite{caulliez98} and figure~7 in \cite{melville_shear_veron_1998}. 

\begin{figure}[h!]
\centering
\includegraphics[]{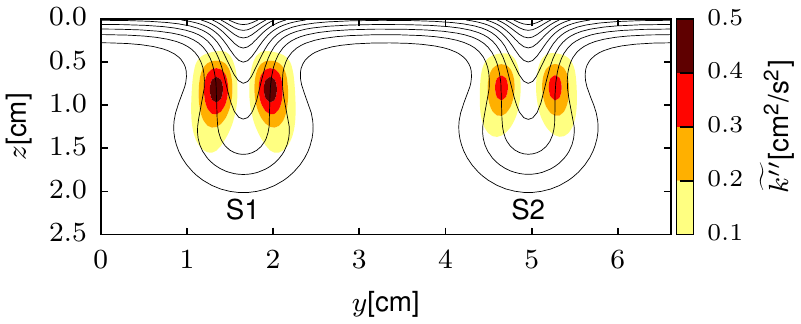}
\caption{\label{fig:crit} Critical layers of the streak instabilities. Filled contours of $\widetilde {k^\dprime}$ are overlaid on line contours of $u$ at $t=24$. }
\end{figure}

\begin{figure}[h!]
\centering
\includegraphics[]{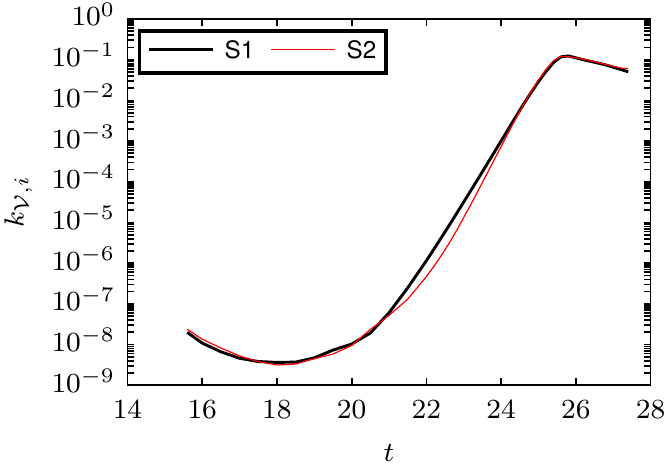}
\caption{\label{fig:kV} Temporal growth of the streak instabilities. }
\end{figure}

The nature of the observed streak instability can be elaborated using a triple decomposition~\cite{onder2020receptivity}:
\begin{equation}
\vec u= \langle \vec u \rangle +\underbrace{\vec  {\widetilde{u^\prime}} + \vec  {u}^\dprime}_{\vec {u}^{\prime}},
\end{equation}
where $\langle \vec u \rangle$ is the plane averaged velocity, $\widetilde{ \vec  u^\prime}$ represents the streamwise-constant fluctuations, i.e., $\partial {\widetilde{ \vec u^\prime}}/\partial x=\vec 0$, and $\vec u^\dprime$ represents the residual fluctuations. When the streaks start to oscillate, the energy of residual fluctuations rapidly rises. This growth of energy can be measured by the streamwise-averaged residual energy,  $\widetilde {k^\dprime}(y,z,t)=\widetilde{u_i^{\prime\prime}u_i^{\prime\prime}}/2$.  Figure~\ref{fig:crit} demonstrates the contours of $\widetilde {k^\dprime}$ at $t=24\mbox{~s}$. The local nature of the instability is clearly visible, where the energy concentrates in the spanwise shear layers at both sides of the streaks at a depth $z\approx 0.8\mbox{~cm}$.  Figure~\ref{fig:kV} further shows the temporal evolution of the length-normalized kinetic energy in each streak subregion,
\begin{equation}
\label{eq:kv}
k_{\mathcal V,i}(t)=\frac{1}{2L_{y,i}}\int\int_{A_i} \widetilde{u_i^\dprime u_i^\dprime} (y,z,t) \mathrm d y \mathrm d z,
\end{equation}
where $L_{y,i}$ is spanwise length of the streaks regions, i.e., half of the spanwise domain length. Starting from $t\approx 21\mbox{~s}$, exponential growth is observed on both streaks. The growth of this modal instability lasts until $t\approx 26\mbox{~s}$.

\section{Conclusions}
The generation, amplification and breakdown of surface streaks in a wind-stress boundary layer are analysed by means of linear non-normal growth theory and direct numerical simulations. The parameters of the flow model are tuned to mimic a previous wind-wave tank experiment in \cite{melville_shear_veron_1998}. A bypass transition scenario similar to that in wall boundary layers is observed, where streamwise vortices amplify streamwise streaks that are located adjacent to the free surface. The streaks grow and penetrate into deeper waters until they develop sinuous instabilities originating from spanwise shear layers. Turbulence quickly sets in following the streak instabilities and slows down the boundary layer.   

\section{Acknowledgements}
A\"O and PLFL ackknowledge a Tier~2 grant from Ministry of Education of Singapore to National University of Singapore. WT is supported by a grant from Ministry of Science and Technology of Taiwan.  The computational work for this article was fully performed on resources of the National Supercomputing Centre, Singapore (https://www.nscc.sg).

\bibliographystyle{afmc}
\bibliography{allpapers.bib}

\end{document}